\documentclass{PoS}
\usepackage{epsfig,graphicx}
\usepackage{amsmath}
\usepackage{amssymb}

\title{Chiral symmetry, di-electrons and charm}

\ShortTitle{Chiral symmetry, di-electrons and charm}

\author{\speaker{Burkhard KAMPFER}%
         \thanks{Also at TU Dresden, 01062 Dresden, Germany.}\\
        FZ Dresden-Rossendorf, 01314 Dresden, Germany\\
        E-mail: \email{kaempfer@fzd.de}}

\author{Thomas HILGER, Henry SCHADE, Robert SCHULZE\\
        FZ Dresden-Rossendorf, 01314 Dresden, Germany\\
        E-mail: \email{t.hilger@fzd.de},
                \email{h.schade@fzd.de},
                \email{r.schulze@fzd.de}}
        
\author{Gyorgy WOLF\\
        KFKI RMKI, H-1525 Budapest, POB 49, Hungary\\
        e-mail: \email{wolf@rmki.kfki.hu}}

\abstract{We survey some prospects of identifying further in-medium modifications
of hadrons in a strongly interacting medium with respect to ongoing experiment series of the
HADES Collaboration and planned experiments of the CBM Collaboration
at FAIR. Di-electrons, strange and charm mesons are considered and their
potential for signaling imprints of chiral restoration is highlighted.}

\FullConference{XLVIII International Winter Meeting on Nuclear Physics 
in Memoriam of Ileana Iori \\
25-29 January 2010\\
Bormio, Italy}

\title{Chiral symmetry, di-electrons and charm} 

\begin{document}

\section{Introduction}

In-medium hadron spectroscopy aims at quantifying the modifications
of spectral functions of hadrons embedded in a strongly interacting
medium. The Brown-Rho scaling prediction (cf.~\cite{Leupold:2009kz,Rapp:2009yu}
for recent surveys on this topic and further references therein)
emphasized the tight relation of hadron masses and the chiral condensate,
however, a strong impact of other condensates is conceivable
(e.g., the poorly known four-quark condensates \cite{Thomas:2005dc,Thomas:2007gx}
or the gluon condensates \cite{Morita:2007hv,Morita:2007pt}).
Due to the direct vector meson ($V$) decays 
$V \to \gamma^* \to e^+ e^-, \mu^+ \mu^-$ a measurement of the spectral
function becomes feasible. Until now, broadening effects of the
$\rho$ meson have been verified in cold nuclear matter \cite{Wood:2008ee}
and in hot nuclear matter, see \cite{Rapp:2009yu}.  
From a measurement of the transparency ratio a strong broadening
of the $\omega$ meson in cold nuclear matter is deduced \cite{:2008xy}.

With respect to the ongoing experiment series of the HADES Collaboration 
and the planned experiments by the Compressed Baryon Matter Collaboration
at FAIR
we report here prospects for identifying medium modifications in the
di-electron and strange/charm meson channels.     

\section{Di-electrons}

\begin{figure}
\centering
\includegraphics[width=0.45\textwidth]{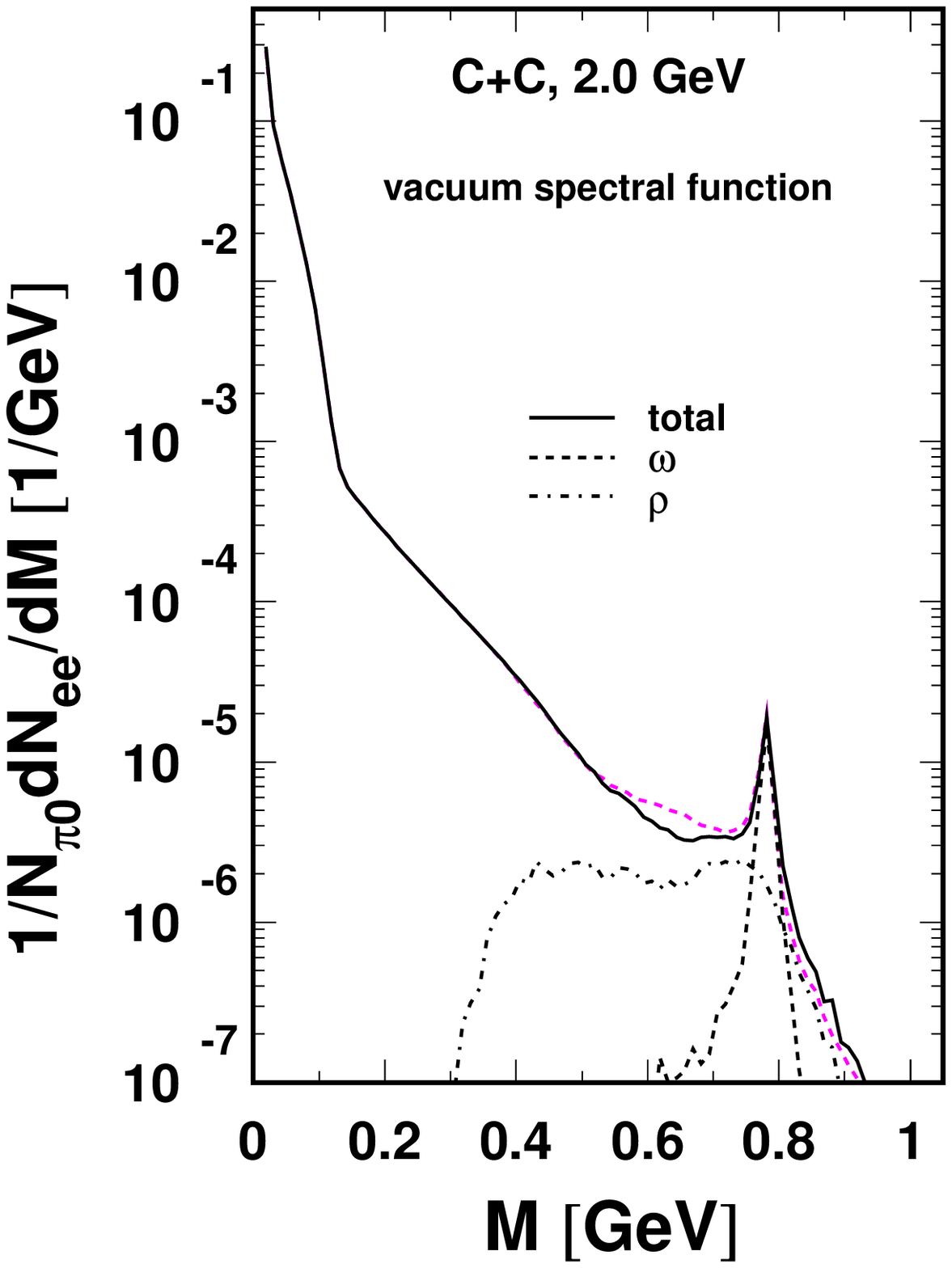}
\includegraphics[width=0.45\textwidth]{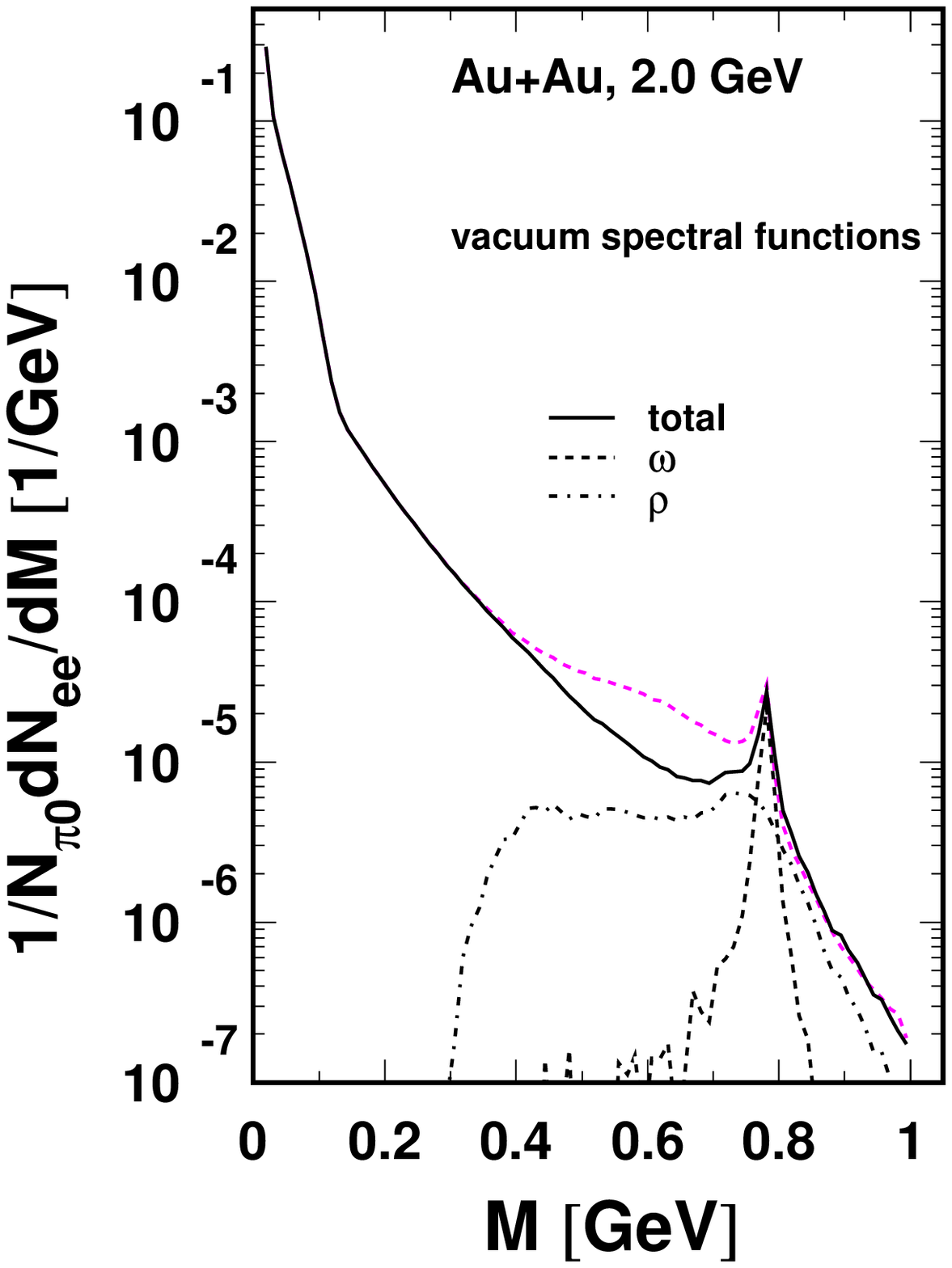}
\caption{Invariant $e^+ e^-$ mass spectra in collisions
C(2 AGeV) + C (left panel) and Au(2 AGeV) + Au (right panel). 
The solid curves employ the vacuum spectral functions of 
$\rho$ (dot-dashed curves) and $\omega$ (dotted curves), while the dashed magenta
curves include in-medium effects. The contributions of $\pi^0$, $\eta$
and $\Delta$ Dalitz decays and bremsstrahlung are not depicted separately.
From \cite{Barz:2009yz}.} 
\label{fig_ee}
\end{figure}

Figure~\ref{fig_ee} (left panel) exhibits the di-electron spectrum
as a function of the invariant mass for collisions C(2 AGeV) + C
(for details of the transport model of BUU type cf.~\cite{Barz:2009yz}). 
The solid black curve is for the superposition of $\pi^0$, $\eta$
and $\Delta$ Dalitz decays and bremsstrahlung as well as the separately displayed
vacuum spectral functions of $\rho$ and $\omega$, while the dashed
magenta curve depicts the modifications for in-medium spectral
functions including collision broadening and extreme mass shifts
by $-100$ MeV and $-50$ MeV
(cf.~\cite{Zschocke:2002mn} for corresponding QCD sum rule evaluations 
and the impact of quark condensates), 
respectively. Despite the assumed very
drastic medium modifications, there is only a tiny imprint in the
spectrum since the small system is too short living, with spectral
functions evolving rapidly to the vacuum ones. 

The prospects for verifying medium modifications are better for
larger nuclei, as depicted in the right panel of Fig.~\ref{fig_ee}
for Au(2 AGeV) + Au. The high-compression stage lasts significantly
longer signaled by the stronger contribution of ''shining'' vector mesons.
There is no indication of a double-hump $\omega$ signal caused by the 
shining in-medium mass-shifted contribution and the final vacuum decays
in both cases considered.

In contrast to the data \cite{Agakichiev:2006tg} for the reaction     
C(2 AGeV) + C, the data \cite{Agakishiev:2007ts} 
for the reaction C(1 AGeV) + C require a strong nucleon-nucleon bremsstrahlung,
as pointed out in \cite{Bratkovskaya:2007jk}. In fact, the estimates
in \cite{Kaptari:2005qz,Kaptari:2009yk} within the one-boson exchange
model point to a strong bremsstrahlung contribution in proton-neutron
collisions at 1 GeV which is over-shined by the $\Delta$ Dalitz
yield at higher beam energies. 

\section{Open strange and charm mesons in nuclear matter}

\begin{figure}[hb]
\centering
\includegraphics[width=0.45\textwidth]{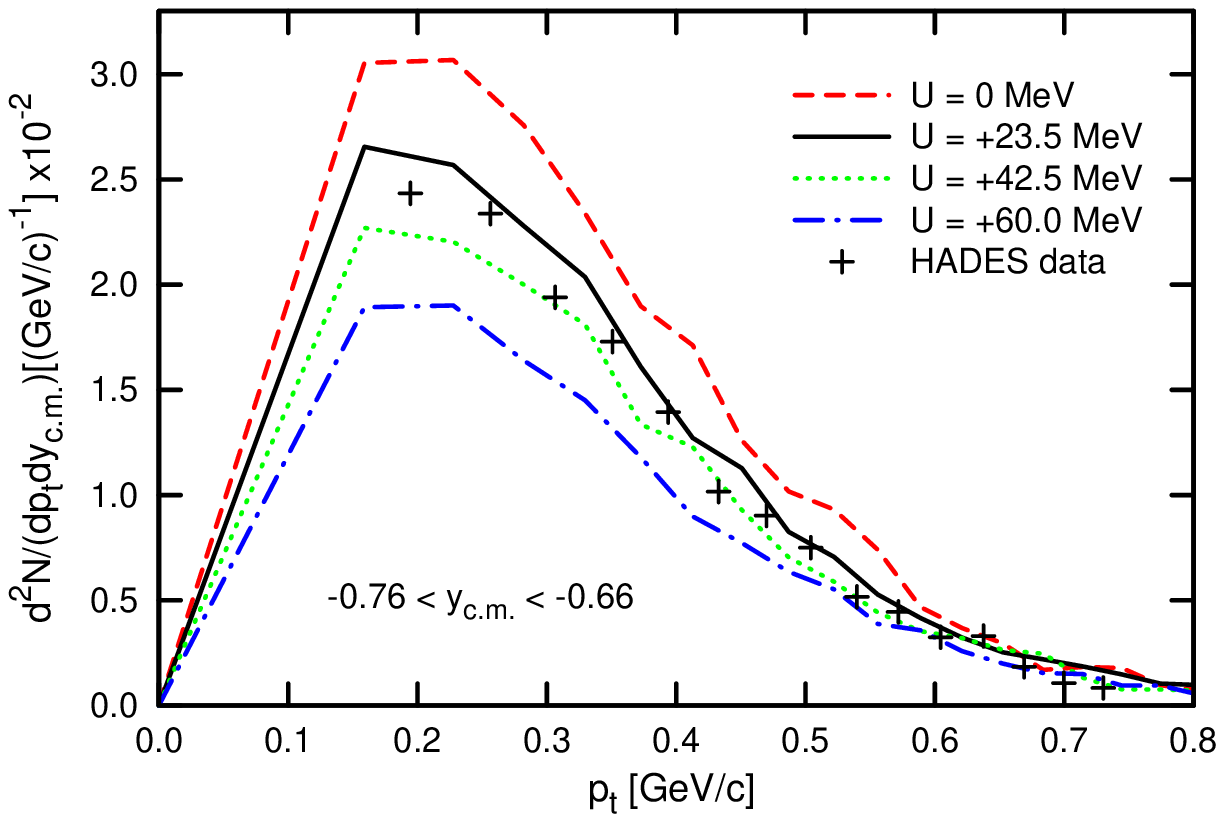} \hfill
\includegraphics[width=0.45\textwidth]{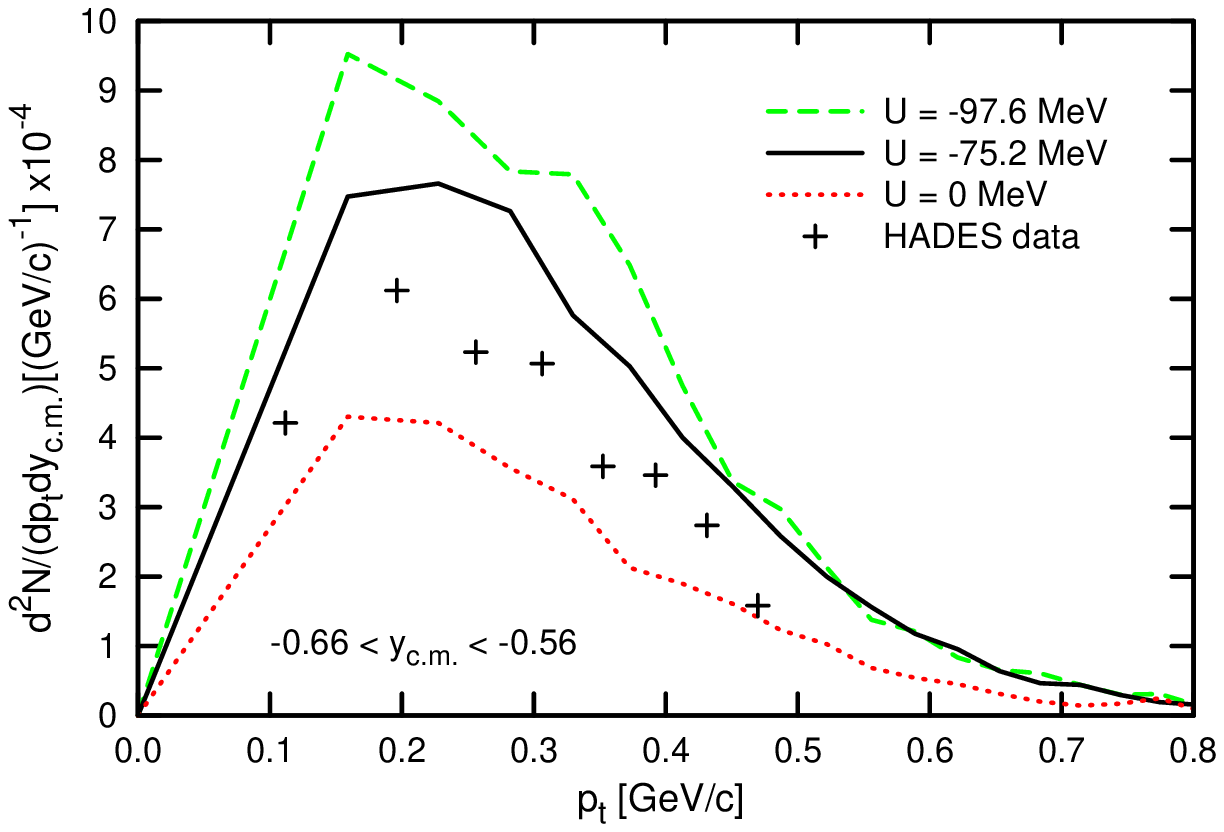}
\caption{Transverse momentum distribution of kaons 
(left panel, rapidity interval $-0.76 < y_{c.m.} < -0.66$) 
and anti-kaons (right panel, rapidity interval $-0.66 < y_{c.m.} < -0.56$) 
for several values of the potential $U$ as indicated by the labels.
All other settings of the BUU model \cite{Schade:2009gg} are frozen in.
Data from \cite{Agakishiev:2009ar}.}
\label{fig_HS}
\end{figure}

In a schematic manner, in-medium modifications of hadrons may be encoded
in effective masses, e.g. parameterized as $m^* = m_0 + U n/n_0$,
where $m_0$ stands for the vacuum mass and the interaction is condensed
into the ''potential'' $U$, which can be related to the real part of the
self-energy. The dependence on the net baryon density $n$
can be involved;  usually the linear density approximation around saturation
density $n_0$ is employed. Figure \ref{fig_HS} exhibits two examples
for the transverse momentum distribution of $K^+$ and $K^-$ in collisions
of Ar(1.756 AGeV) + KCl calculated in a transport model of BUU type,
cf.~\cite{Schade:2009gg}. Clearly, an attractive (repulsive) net interaction with 
surrounding nucleons translates into a negative (positive) potential
which enhance (deplete) the respective anti-kaon (kaon) abundances and
modify the phase space distribution as seen in Fig.~\ref{fig_HS}. 
Precision data for transverse momenta $p_t \sim 200$ MeV/c in
various rapidity bins, in particular at mid-rapidity, are needed to pin down reliably 
the actual values of the potentials.    

\begin{figure}[ht]
\vskip -3mm
\includegraphics[width=0.49\textwidth]{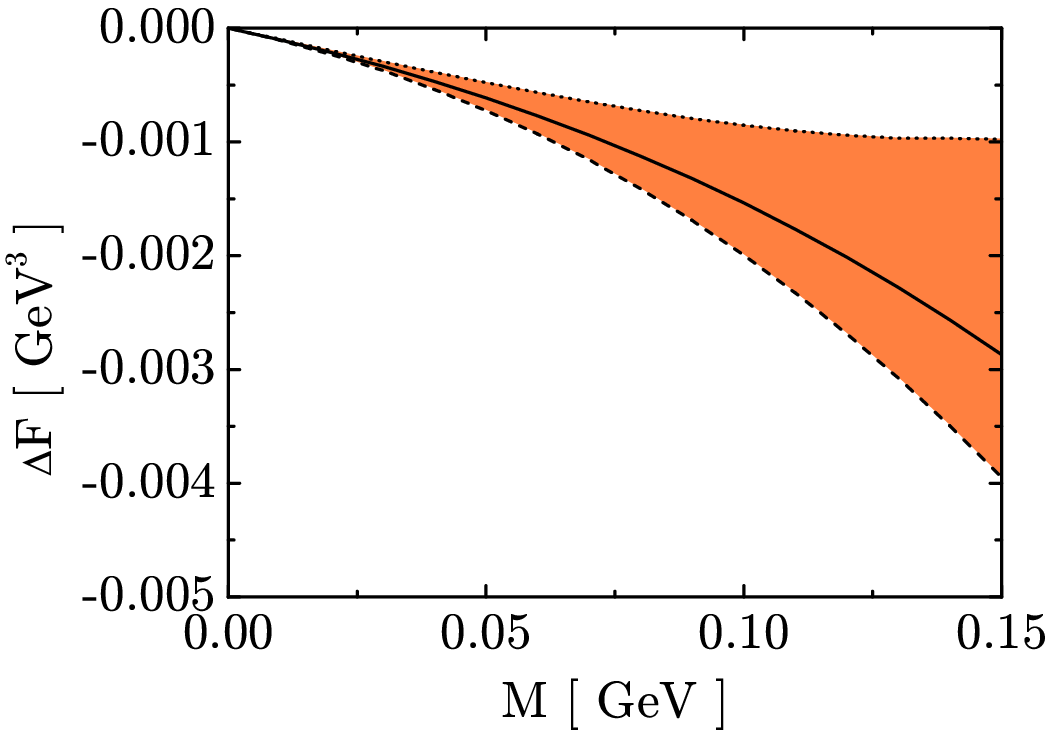}
\includegraphics[width=0.49\textwidth]{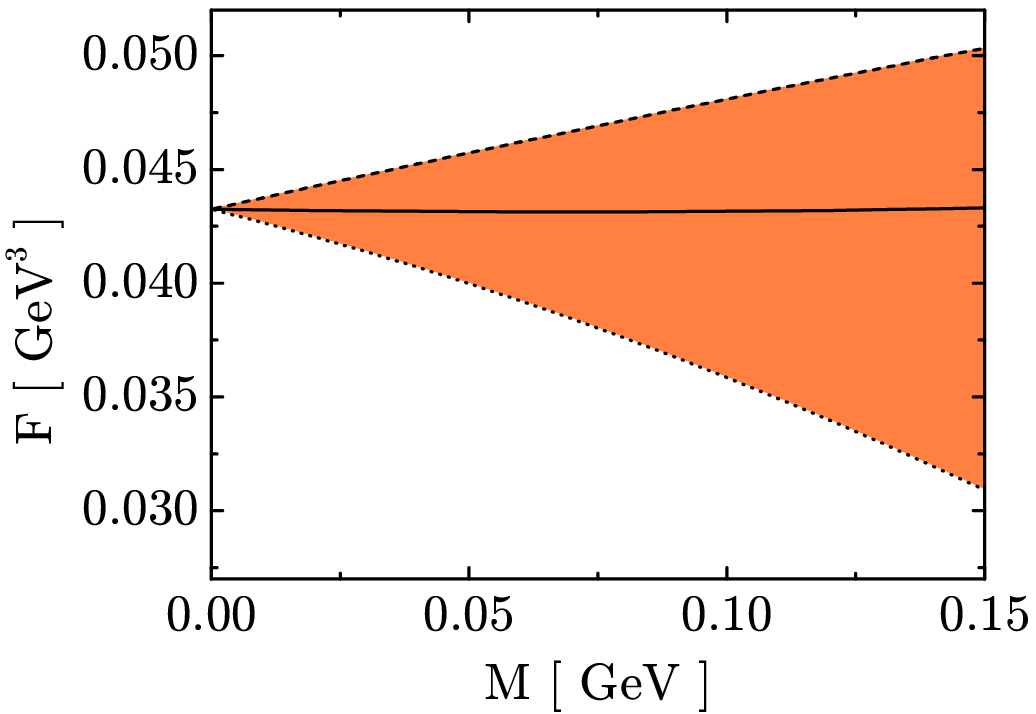}
\caption{The quantities $\Delta F$ (left panel) and $F$ (right panel)
entering the QCD sum rule for $D$ mesons.
Line codes as in  \cite{Hilger:2008jg}.  }
\label{fig_TH}
\end{figure}

The calculation of $K^\pm$ spectral functions, from which the potential
$U$ can be deduced, may be performed within hadronic interaction approaches,
such as in \cite{Lutz:1997wt,Lutz:2007bh}, or may rely on QCD sum rules. 
The evaluation for $D^\pm$ mesons, for example, 
is based on the current-current correlator
$\Pi(q) = i \int d^4x \, e^{iqx} \langle \langle \, {\cal T}
\left[j(x) j^\dagger(0)\right] \rangle \rangle$
with $j(x) = i \bar{d} \gamma_5 c$ being the pseudo-scalar flavor changing current, 
${\cal T}$ the time ordering operator, and $\langle \langle \cdots \rangle \rangle$ denotes 
the Gibbs average.
Applying the operator-product expansion (OPE) at large squared momenta $q_0^2 = -Q^2$
relates the correlator to QCD condensates.
On the other hand, using a Borel transformed in-medium dispersion relation 
in the complex energy plane maps the even ($e$) and the odd ($o$) part 
of the OPE to moments of the spectral density $\Delta \Pi(s)$
\begin{eqnarray} \label{eq:mom_sys}
e &\equiv&
\int_{s_0^-}^{s_0^+} ds \, s \, \Delta \Pi {\rm e}^{-s^2 /M^2} 
= m_+ F_+ {\rm e}^{-m_+^2 / M^2}
+ m_- F_- {\rm e}^{-m_-^2 / M^2}, \label{mom1}\\
o &\equiv&
\int_{s_0^-}^{s_0^+} ds \, \Delta \Pi {\rm e}^{-s^2 /M^2} = F_+ {\rm e}^{-m_+^2 / M^2} -
F_- {\rm e}^{-m_-^2 / M^2} \label{mom2} \: ,
\end{eqnarray}
where we employed a pole ansatz
$\Delta \Pi (s) = \pi F_+ \, \delta (s - m_+) - \pi F_- \, \delta(s + m_-)$,
which separates the perturbative continuum from the low-lying states 
by the threshold parameters $s_0^\pm$; $M$ is the Borel parameter.
For $m_\pm = m \pm \Delta m$ and $F_\pm = F \pm \Delta F$ being independent 
of the Borel mass, $F_\pm$ can be eliminated, 
and the mass splitting and mass centroid of $D^-$ and $D^+$ are given by
\begin{equation} \label{eq:m_sys}
    \Delta m = \frac12 \frac{ o e^\prime - e o^\prime}{ e^2 + o o^\prime } \: , \quad
    m = \sqrt{ \Delta m^2 - \frac{ e e^\prime + \left( o^\prime \, \right)^2 }
    { e^2 + o o^\prime } } \: .
\end{equation}
The strengths can then be obtained by
$F_\pm = e^{(m \pm \Delta m)^2/M^2} [e \pm (m \mp \Delta m)o]/(2m)$;
$F$ and $\Delta F$ can be written in terms of the mass splitting and mass centroid as
\begin{eqnarray} \label{eq:F_sys}
    \Delta F &=& \frac12 \frac{ {\rm e}^{(m^2+\Delta m^2)/M^2}}{m}
    \left[ (e-o\Delta m)\,{\rm sinh}\left(\frac{2m\Delta m}{M^2}\right) + 
    o m \,{\rm cosh} \left(\frac{2m\Delta m}{M^2}\right) \right]
    \: ,
    \label{eq:delta_F}
    \\
    F &=& \frac12 \frac{ {\rm e}^{(m^2+\Delta m^2)/M^2}}{m}
    \left[ (e-o\Delta m)\,{\rm cosh}\left(\frac{2m\Delta m}{M^2}\right) + 
    o m \,{\rm sinh} \left(\frac{2m\Delta m}{M^2}\right) \right]
    \label{eq:F_cms}
\end{eqnarray}
which are exhibited in Fig.~\ref{fig_TH}. 
One can show that $F_\pm(M_0)$ are extremal if $m_\pm(M_0)$ are extremal,
which allows for a consistent evaluation reported in \cite{Hilger:2008jg}. 
The found splitting
of masses of $D^+$ and $D^-$ at non-zero baryon density corresponds
to the splitting of $K^-$ and $K^+$. It should be emphasized that up to
mass-dimension six the density dependence of the condensates
$m_c \langle \bar d d \rangle$, 
$\langle d^\dagger d \rangle$,
$\langle d^\dagger D_0^2 d \rangle$, and
$\langle \frac{\alpha_s}{\pi} G^2 \rangle$
drive the in-medium change of the $D$ mesons. As pointed out in \cite{Hilger:2010zb},
the chiral condensate $\langle \bar d d \rangle$ in combination with 
the charm quark mass $m_c$ is the leading term for difference sum rules
of chiral partners with the quark structure $qQ$, i.e. a light quark ($q$) 
and a heavy quark ($Q$), such as for $D$ meson states.
That means, the difference of spectral moments, e.g.\ in the
pseudo-scalar -- scalar or vector -- axial-vector channels,
is diminished with dropping modulus of the chiral condensate
which may serve as order parameter of chiral restoration. 

\section{Condensates at high temperature}

To employ QCD sum rules near the deconfinement transition one has to
go beyond the low-density and/or low-temperature expansion.
Following \cite{Morita:2007hv,Morita:2007pt}
the gluon condensate can be derived from the interaction measure $e-3p$,
determined by the QCD trace anomaly,
and the equation of state, which relates the energy density $e$ and
the pressure $p$, as
\begin{eqnarray}
&&\left<\frac{\alpha_{s}}{\pi}G^{2}\right> = \left<\frac{\alpha_{s}}{\pi}G^{2}
\right>_{0}-\frac{8}{9}(e-3p),
\label{cond1}\\
&&\left<\frac{\alpha_{s}}{\pi}\left((uG)^{2}-\frac{G^{2}}{4}\right)\right> 
 =  -\frac{3}{4}\frac{\alpha_{s}}{\pi}(e+p),
\label{cond2}
\end{eqnarray}
where the subscript $0$ refers to the vacuum value. Contributions from light quarks
to (\ref{cond1}) are omitted in order to focus on the continuation
to finite net baryon densities.

The presently most trusted values for the interaction measure and the equation
of state at vanishing density (or equivalently quark chemical potential
$\mu)$ are results of Monte Carlo lattice simulations \cite{Bazavov:2009zn}.
They can be extrapolated to finite chemical potential using a self-consistent
quasi-particle model \cite{Schulze:2007ac,Schulze:2008de}.
\begin{figure}[hb]
\includegraphics[width=0.45\textwidth]{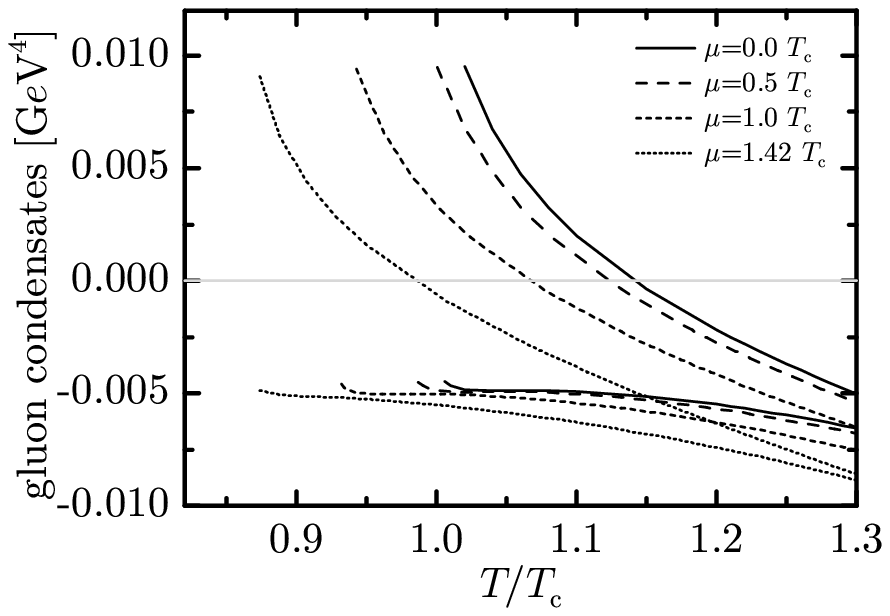} \hfill
\includegraphics[width=0.45\textwidth]{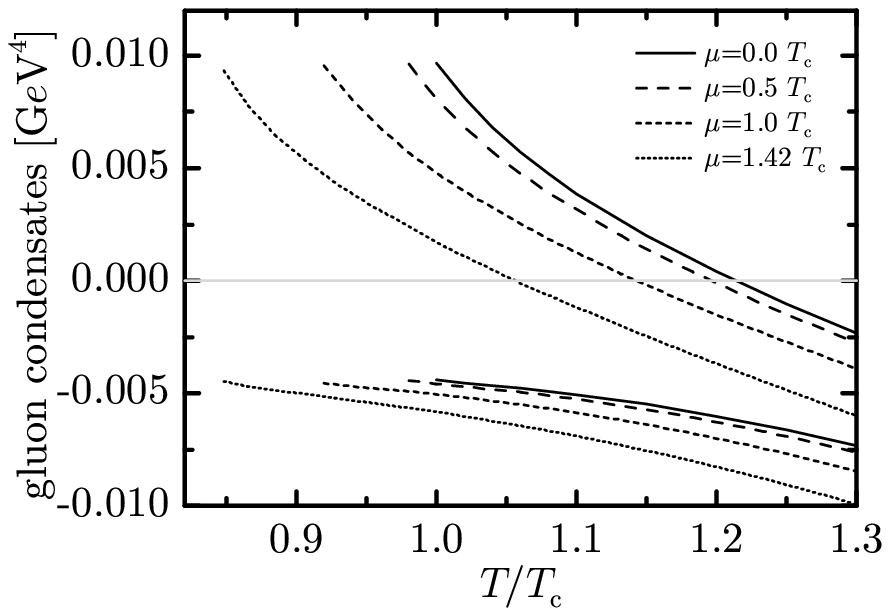}
\caption{Gluon condensates (\protect\ref{cond1}) 
(upper curves) and  (\protect\ref{cond2})
(lower curves) as a function of the temperature $T$ for
several values of the quark chemical potential $\mu$
(see keys). The employed
quasi-particle model is adjusted to recent lattice calculations
\cite{Bazavov:2009zn}. The left (right) panel shows results
for lattice temporal extent $N_{\tau}$ = 6 (8).
The critical temperature has been fixed to $T_c$= 190 MeV.}
\label{fig_RS}
\end{figure}

There is still some uncertainty about the continuum extrapolation
of such lattice calculations. The current results were obtained on
lattices with temporal extent $N_\tau$ of 6 and 8. Adjusting the
quasi-particle model to both cases and employing the resulting values
for $e$ and $p$ give the gluon condensates as exhibited in Fig.~\ref{fig_RS}.
Both condensates are monotonically decreasing functions of the temperature
$T$. All curves start at the presumed locus of the phase transition,
i.e. the boundary of the quasi-particle model determined by the
characteristic curve (cf.~\cite{Schulze:2007ac})
emerging from $T=T_c$ at $\mu=0$.
Numerically, a scaling law
$X \left( \mu/T_c = T(\mu=0)/T_c - 0.045 T/T_c - 0.027 (T/T_c)^2 \right) = const$
is observed
for $X$ being one of the condensates (\ref{cond1}) and (\ref{cond2}).
Additionally, both gluon condensates have a steeper slope
for smaller $N_{\tau}$ with the difference being about the same size
as the difference in the interaction measure of the lattice calculation
leading to the assumption that future improved lattice results might lead to
flatter curves. The condensates considered here determine, to a large extent,
the behavior of $J/\psi$ near the deconfinement region.

\section{Summary}

In summary we emphasize that (i) large systems are required
to enhance chances for verifying vector meson modifications
in the di-electron channel in heavy-ion collisions at SIS18
energies. (ii) Phase space distributions of kaons and anti-kaons
are sensitive to mass shifts encoded in interaction potentials. 
(iii) $D$ mesons exhibit a strong
coupling to the chiral condensate and make them useful probes
of chiral symmetry restoration imprints. (iv) A few QCD condensates
are accessible at temperatures and baryon densities in the
deconfinement transition region.   

Acknowledgments: The work is supported by T71989 (Hungary),
BMBF 06DR9059 and GSI-FE (Germany).

\end{document}